\documentclass[useAMS,usenatbib]{mn2e}
\usepackage{float}
\usepackage{wrapfig}
\usepackage{graphicx}
\usepackage{tabularx}
\usepackage{subfig}
\usepackage{amsmath}

\usepackage{amsthm}
\usepackage{amssymb}
\usepackage{textcomp, gensymb}

\def \OIII {[O\,{\sc iii}]$\lambda$5007\,\AA}
\def \NII {[N\,{\sc ii}]$\lambda$6584\,\AA}
\def \ha {H$\alpha$}

\def \Ba {Ba\,{\sc ii}}
\def \teff {\ifmmode{t_{{\rm eff}}}\else{$t_{{\rm eff}}$}\fi}
\def \texp {\ifmmode{t_{{\rm exp}}}\else{$t_{{\rm exp}}$}\fi}
\def \vhel{\ifmmode{V_{{\rm hel}}}\else{$V_{{\rm hel}}$}\fi}
\def \vexp {\ifmmode{V_{{\rm exp}}}\else{$V_{{\rm exp}}$}\fi}
\def \vsys {\ifmmode{V_{{\rm sys}}}\else{$V_{{\rm sys}}$}\fi}

\def\msun{\ifmmode{{\rm\ M}_\odot}\else{${\rm\ M}_\odot$}\fi}
\def\rsun{\ifmmode{{\rm\ R}_\odot}\else{${\rm\ R}_\odot$}\fi}

\def\myr{\ifmmode{{\rm\ M}_\odot{\rm\ yr}^{-1}}
         \else{${\rm\ M}_\odot$ yr$^{-1}$}\fi}

\def\kms{\ifmmode{~{\rm km\,s}^{-1}}\else{~km s$^{-1}$}\fi}

\begin{document}

\title[The Planetary Nebula LoTr~1]{Two rings but no fellowship: LoTr~1 and its relation to planetary nebulae possessing barium central stars.}

\author[A.A. Tyndall et al.]{A.A.~Tyndall$^{1,2}$\thanks{E-mail: amy.tyndall@postgrad.manchester.ac.uk}, D. Jones$^{2}$, H.M.J.~Boffin$^{2}$, B. Miszalski$^{3,4}$, F. Faedi$^{5}$, M. Lloyd$^{1}$, 
\newauthor 
J.A. L\'{o}pez$^{6}$, S. Martell$^{7}$, D. Pollacco$^{5}$, and M. Santander-Garc\'\i a$^{8}$\\
\\
$^{1}$Jodrell Bank Centre for Astrophysics, School of Physics and Astronomy, University of Manchester, M13 9PL, UK \\
$^{2}$European Southern Observatory, Alonso de C\'ordova 3107, Casilla 19001, Santiago, Chile \\
$^{3}$South African Astronomical Observatory, PO Box 9, Observatory 7935, South Africa\\
$^{4}$Southern African Large Telescope. PO Box 9, Observatory 7935, South Africa\\
$^{5}$Department of Physics, University of Warwick, CV4 7AL, UK \\
$^{6}$Instituto de Astronom\'{i}a, Universidad Nacional Aut\'{o}noma de M\'{e}xico, Ensenada, Baja California, C.P. 22800, Mexico\\
$^{7}$Australian Astronomical Observatory, North Ryde, 2109 NSW, Australia\\
$^{8}$Observatorio Astron\'omico National, Madrid, and Centro de Astrobiolog\'\i a, CSIC-INTA, Spain\\
}
\date{Accepted xxxx xxxxxxxx xx. Received xxxx xxxxxxxx xx; in original form xxxx xxxxxxxx xx}

\pagerange{\pageref{firstpage}--\pageref{lastpage}} \pubyear{2013}

\maketitle

\label{firstpage}

\begin{abstract}

LoTr~1 is a planetary nebula thought to contain an intermediate-period binary central star system ( that is, a system with an orbital period, P, between 100 and, say, 1500 days). The system shows the signature of a K-type, rapidly rotating giant, and most likely constitutes an accretion-induced post-mass transfer system similar to other PNe such as LoTr~5, WeBo~1 and A70. Such systems represent rare opportunities to further the investigation into the formation of barium stars and intermediate period post-AGB systems -- a formation process still far from being understood. Here, we present the first detailed analyses of both the central star system and the surrounding nebula of LoTr~1 using a combination of spectra obtained with VLT-FORS2, AAT-UCLES and NTT-EMMI, as well as SuperWASP photometry.
 We confirm the binary nature of the central star of LoTr~1 that consists of a K1~III giant and a hot white dwarf. The cool giant does not present any sign of s-process enhancement but is shown to have a rotation period of 6.4 days, which is a possible sign of mass accretion. LoTr~1 also presents broad double-peaked H$\alpha$ emission lines, whose origin is still unclear. The nebula of LoTr~1 consists in two slightly elongated shells, with ages of 17 000 and 35 000 years, respectively, and with different orientations. As such, LoTr~1 present a very different nebular morphology than A70 and WeBo~1, which may be an indication of difference in the mass transfer episodes. 
 \end{abstract}


\section{Introduction}

\begin{figure*}
\centering
\includegraphics[scale=0.9]{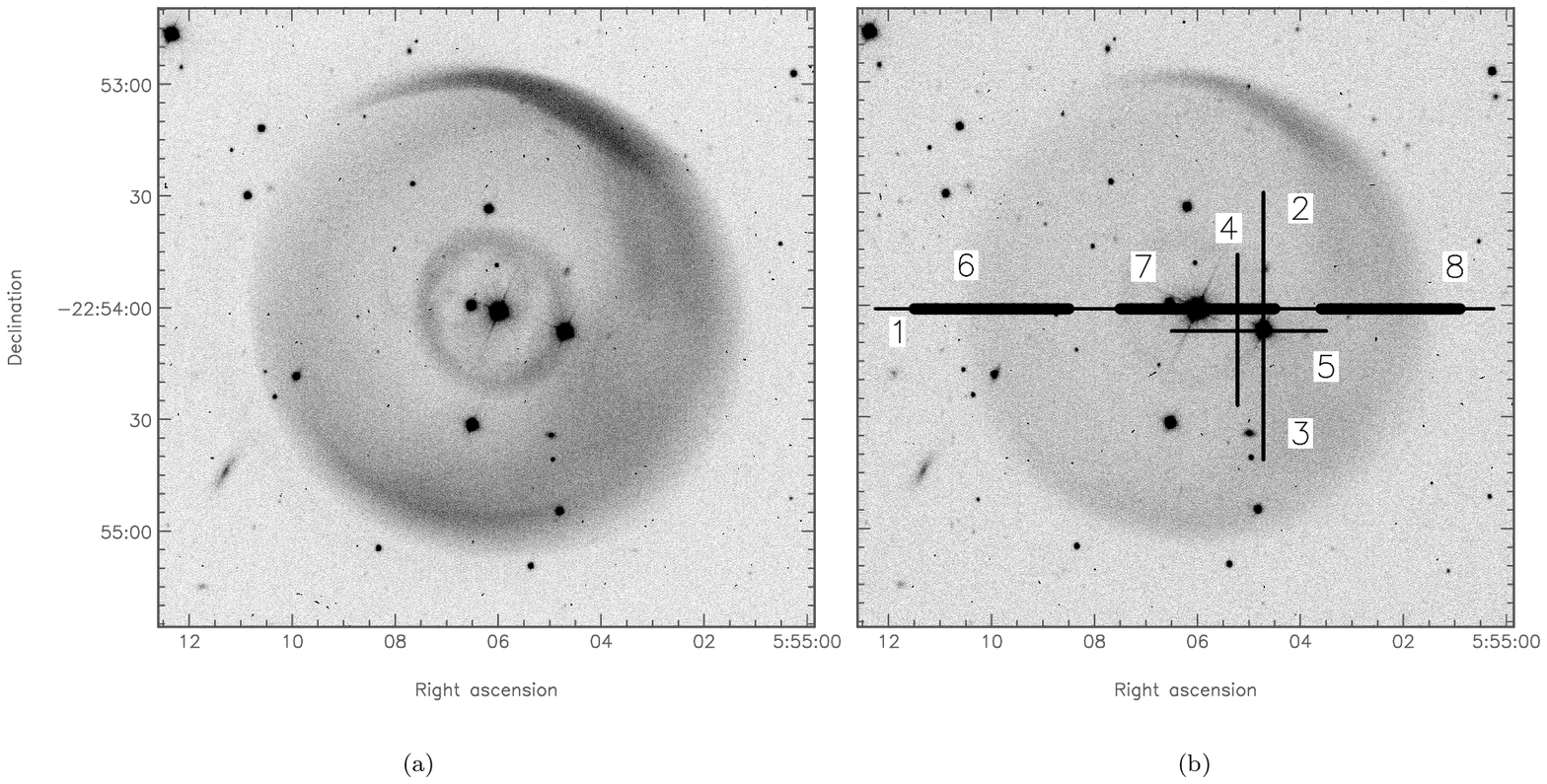}
\caption{Deep, narrowband images of PN LoTr~1, in (a) \OIII{}, and (b) \ha+\NII{}. North is to the top of the image, East is left. The central star is visible at $\alpha$ = 05:55:06.6, $\delta$ = $-$22:54:02.4. Overlaid on image (b) are the slit positions of the spatio-kinematic data presented in section \ref{subsubsec:neb_spec}. Slit 1 (E-W) was acquired using NTT-EMMI and is centered on the central star. Slits 2--4 (N-S) and 5--8 (E-W, width of slits 6, 7 and 8 exaggerated to clearly show the positioning due to overlapping with slit 1) were acquired using AAT-UCLES. All slits were taken in [O\,{\sc iii}].}
\label{fig:lotr1}
\end{figure*}

The interaction between the progenitor star and a binary or planetary companion is believed to shape the resulting planetary nebula (PN), and in some cases even thought to be almost essential for a PN to form \citep{moe06}. The shaping influence of a common-envelope evolution has been studied extensively (see e.g.\ \citealp{tyndall12}, \citealp{jones10a}), and understood in terms of either a collimated fast wind (CFW) carving out an axisymmetric nebula \citep{soker00} or the ejected common-envelope (CE) forming an equatorial density enhancement \citep{nordhaus06} as required by the `Generalised Interacting Stellar Winds' model \citep{gisw}. However, very little is known about intermediate-period (P=100-1500 days, \citealp{vanwinkel07}) post-Asymptotic Giant Branch (AGB) binaries, including their effect on PNe formation and morphology, due to lack of observations. The intermediate period binaries fall between post-CE systems \citep{miszalski09} and visually resolved systems (e.g. \citealp{ciardullo99}). \citet{soker97} claimed that these interacting systems are the most likely CSPNe (central stars of planetary nebulae) to form the classical `butterfly', or bipolar, morphologies, but with few systems known and limited investigations, this has yet to be confirmed. Only by finding and studying CSPNe with intermediate periods can we substantiate this claim and relate the processes at work in formation of PNe by both post-CE and intermediate-period CSPNe. 


The planetary nebula (PN) LoTr~1 ($\alpha$ = 05:55:06.6, $\delta$ = $-$22:54:02.4, J2000) was first discovered by A.J.~Longmore and S.B.~Tritton with the UK 1.2-m Schmidt telescope \citep{longmore80}. It is generally noted that LoTr~1 belongs to the so-called `Abell~35-type' group \citep{bond93} of PNe showing evidence of a binary central star system consisting of a cool central star (a rapidly rotating subgiant or giant), and an optically faint hot companion (a white dwarf with effective temperature, \teff{} $\sim$ 100~kK) since these giant stars are too cool to ionise the surrounding nebula. Four PNe fell into this category: Abell~35 (hereafter, A35), LoTr~5 \citep{thevenin97}, WeBo~1 \citep{bond03} and Abell~70 (\citealp{miszalski12}, hereafter, A70). However, \citet{frew08} determined that A35 is most likely not a true PN, but rather a Str\"omgren zone in the ambient interstellar medium (ISM). This claim is substantiated by \citet{ziegler12}, who find that the central star may in fact have evolved directly from the Extended Horizontal Branch to the White Dwarf (WD) phase (a so-called AGB-manqu\'e star). As such, we choose not to consider Abell~35 in our comparisons among this group.

Another common factor amongst this particular group of PNe is evidence for the existence of `Barium (\Ba{}) stars' \citep{bidelman51} -- population \textsc{i} G/K-type AGB stars that show an over-abundance of carbon and s-process elements, in particular barium \citep{thevenin97}. A now-canonical model for the formation of these \Ba{} stars states that they form not through CE evolution\footnote{We note, however, the existence of some exceptional systems that also experience similar enrichment in close binaries, i.e. the Necklace nebula \citep{miszbofcor13}, and which are most likely linked to dwarf carbon stars.}, as is the case for close binaries found within PNe, but rather via a wind-accretion scenario \citep{boffin88}. Here, the future \Ba{} star is polluted whilst on the main sequence by the wind of its companion (\citealt{luck91}; the companion having dredged up these s-process elements during its thermally pulsing AGB phase), but with the system remaining detached. After the envelope is ejected to form the surrounding nebula, the AGB star evolves into a WD, while the contaminated star retains its chemical peculiarities to form the remnant \Ba{} star. 

One important prediction to come out of the wind-accretion model is that the accreting star, i.e. the future \Ba{} star, also accretes angular momentum from the companion to become a rapid rotator \citep{jeffries96, theuns96}. Indeed, photometric monitoring of LoTr~5 \citep{thevenin97} and Webo~1 \citep{bond03} has revealed that their cool components are in fact rapid rotators with a rotation period of a few days, thus providing further evidence for this formation scenario. This is further evidenced by the fact that \citet{montez10} found that the x-ray emission from the binary central stars of LoTr 5 is most likely due to the chromospheric activity from a spun-up companion.

Most \Ba{} stars are not observed to be within PNe, almost certainly because the lifetime of the PN is very short with respect to the lifetime of the stellar system. However, this does not completely rule out the possibility that some \Ba{} stars may be formed without passing through a PN phase. Indeed, there are a few examples of field stars that have been shown to consist of a rapdidly rotating cool star linked to an optically faint hot component in a similar fashion to the Abell-35 group, but without current evidence for a surrounding PN. 
As mentioned by \citet{bond03}, HD~128220 (P = 872 days) is one such system, made up of an O subdwarf and a G0 giant companion. O subdwarfs overlap with CSPN in terms of their log~g and \teff{}, implying that they too are found to be in a post-AGB phase of evolution \citep{howarth90}. 56 Pegasi is another example, whereby it possess a system consisting of a K0 giant and a hot white dwarf companion with evidence for an overabundance of \Ba{}, and has an estimated orbital period of 111 days \citep{griffin06}. It is possible that both of these systems \emph{have} gone through a PN phase in the past, but it has since dissipated into the surrounding ISM. This fact, as well as the knowledge that in \Ba{} central stars the chemical pollution process happened very recently -- either during or immediately prior to the formation of the PN -- makes it highly important to study such systems as A70 and WeBo~1, as it will allow us to gain greater insight into both the s-process within AGB stars and mass-transfer mechanisms.

In this paper, we present photometric and spectroscopic observations of LoTr~1 and its central star along with complementary data of A70 and WeBo~1, in order to try to relate the evolutionary processes of these three systems\footnote{PN K~1-6 \citep{frew11} is another potential candidate for this group of objects, as it also shows evidence of possessing both a G- or K-type giant (inferred from imagery, looking at both optical and 2MASS near-IR colours) and a very hot sub dwarf or white dwarf (inferred from GALEX archival images) at its core. However, no stellar spectroscopy is available as yet to look for signs of chemical enrichment \citep{frew11}. More recently, Miszalski et al.~(in press) presented evidence for a carbon and s-process enriched giant at the centre of the planetary nebula Hen~2-39. Due to it being a newly investigated system, Hen~2-39 is not included in this study either.}, and see if they belong to a common ``fellowship'' of s-process enriched cool CSPN inside ring-like nebulae.

\begin{figure*}
\centering
\includegraphics{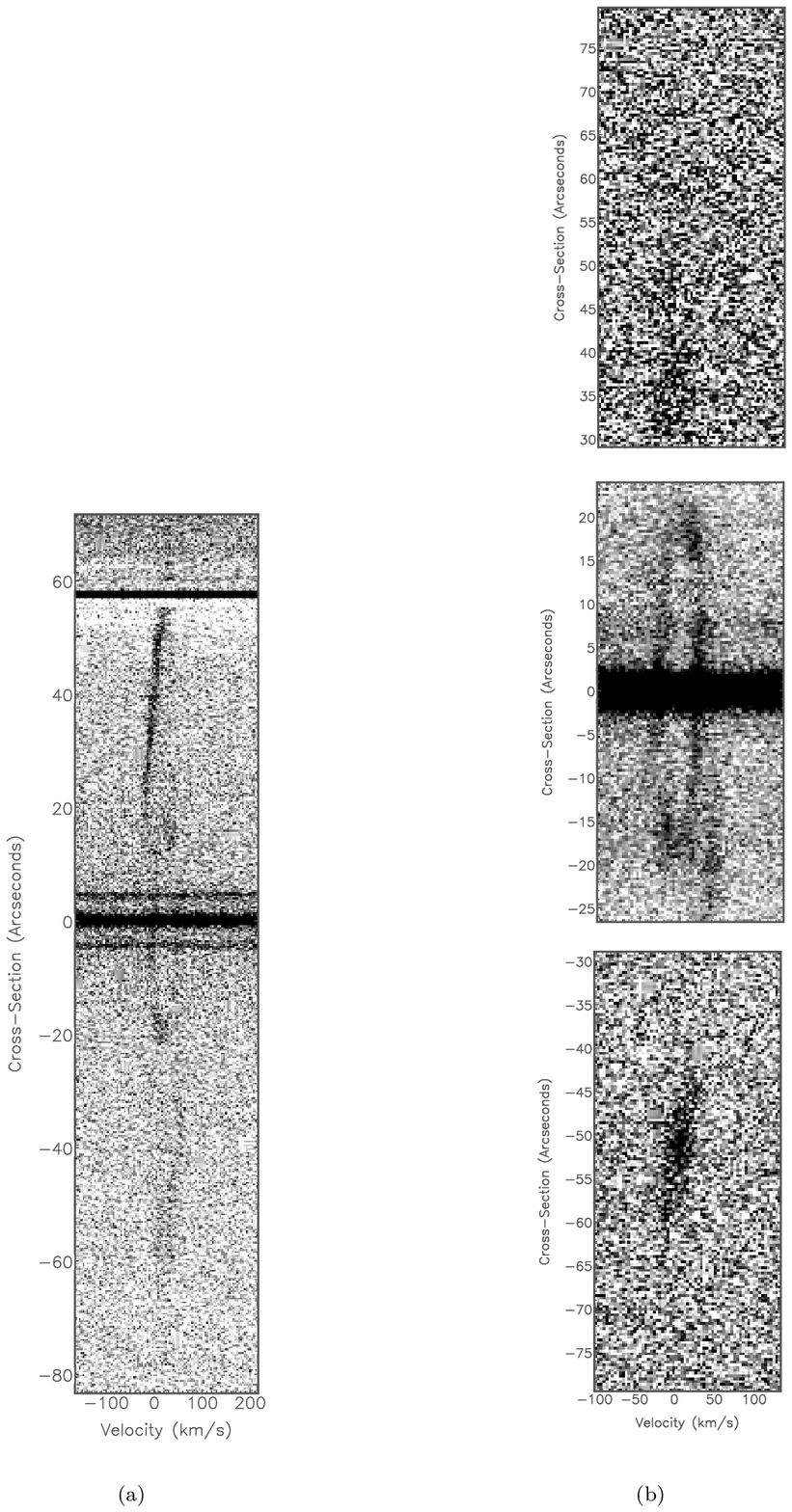}
\caption{PV arrays showing reduced, longslit \OIII{} NTT-EMMI and AAT-UCLES spectra of LoTr~1, both aligned E-W, to show the overall nebular structure. West is to the top of the array. The velocity axis is heliocentric velocity, \vhel{}. The display scale has been modified to highlight the spatio-kinematic features referred to in the text. Cross-section 0\arcsec{} defines where the central star is found. The gap between the two CCDs in figure (a) is visible as a black strip across the frame at cross-section 58\arcsec{} (see text for details).}
\label{fig:lotr1_full_slits}
\end{figure*}

\begin{figure*}
\centering
\includegraphics[scale=0.9]{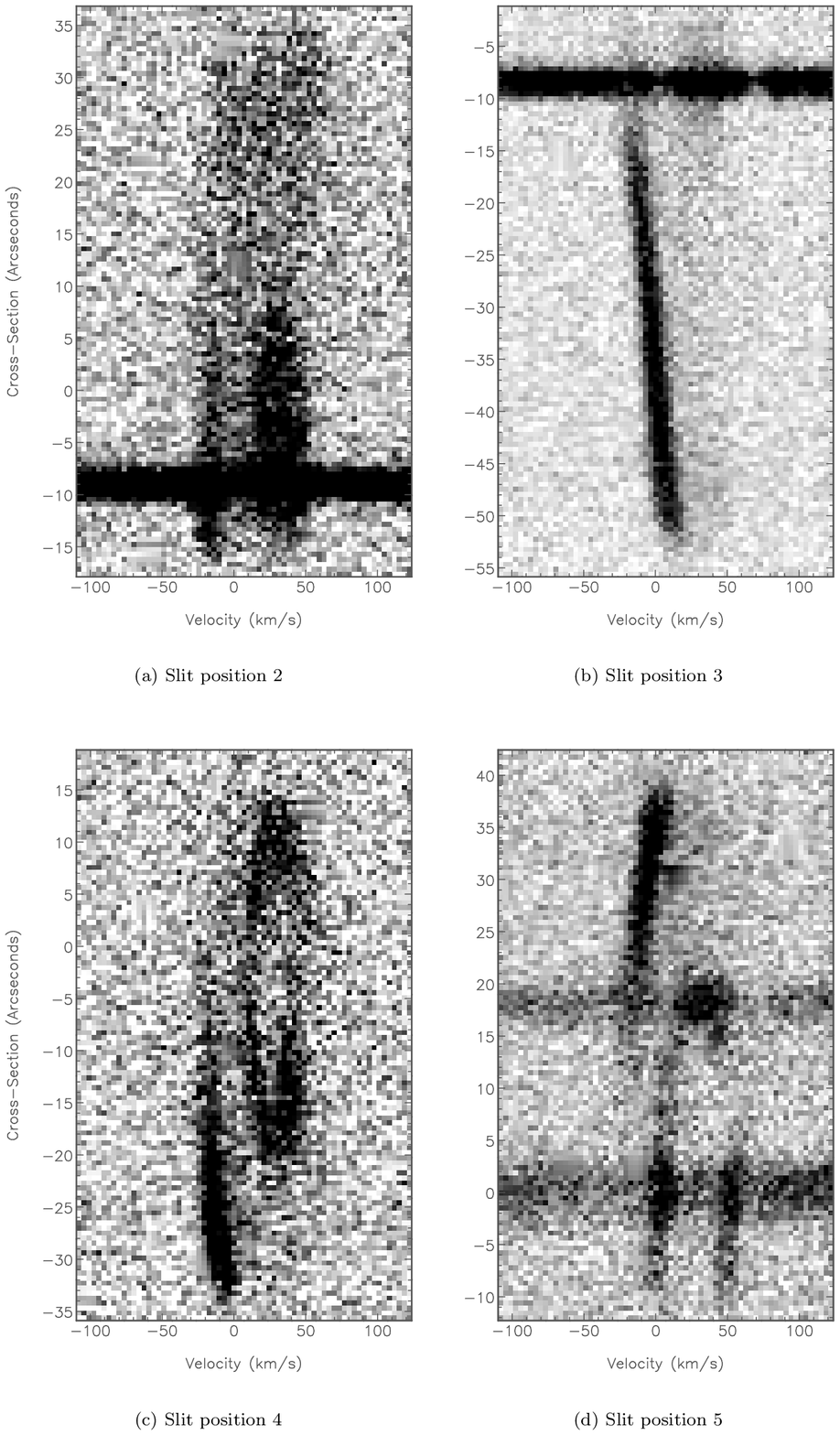}
\caption{PV arrays showing reduced \OIII{} AAT-UCLES spectra of LoTr~1. The velocity axis is heliocentric velocity, \vhel{}. The display scale has been modified to highlight the spatio-kinematic features referred to in the text. Slits 2--4 are N-S (North is to the top of the array), slit 5 is E-W (West is to the top of the array). Cross-section 0\arcsec{} defines where the central star is found. The  continuum of a field star is visible at cross-section $-$8\arcsec{} in figures (a) and (b), and at cross-section +18\arcsec{} in figure (d).}
\label{fig:lotr1_aat_1}
\end{figure*}

\section{Observations and Analysis}
 \label{sec:lotr1}

	\subsection{LoTr1}
	 \label{subsec:lotr1_obs}

\subsubsection{Imaging}
  \label{subsubsec:lotr1_image}

The deep \OIII{} and \ha + \NII{} images shown in figures \ref{fig:lotr1}(a) and (b) were acquired on 2005 March 03 using the red arm of the European Southern Observatory (ESO) Multi-Mode Instrument (EMMI; \citealt{dekker86}), mounted on the 3.6m New Technology Telescope (NTT) of the La Silla Observatory. EMMI was used with the mosaic of two MIT/LL CCDs, 2048 $\times$ 4096 15 $\mu$m pixels. The exposure time, \texp{}, in each filter was 1800s and the binning set to $2\times2$ ($\equiv$ 0.33\arcsec{} pixel$^{-1}$). The seeing was $\sim$0.7\arcsec{}.

 The images show LoTr~1 to have an apparent double-shell structure, with the central shell having a circular profile with an angular diameter of 47\arcsec{}$\pm$4\arcsec{}, and the outer shell a more irregular, but still roughly circular, appearance, with a diameter of 2\arcmin{}26\arcsec{}$\pm$4\arcsec{}. Furthermore, in \OIII{} the outer shell appears brighter in the northwest and southeast -- this, coupled with the slight deviation from circular symmetry, could be considered evidence for an inclined, elongated structure, where the brighter areas result from a projection effect.

Interaction with the ISM is also frequently invoked to explain deviation from symmetry and brightening of the nebular shell (e.g., \citealt{jones10a}), however this would not produce the apparent axisymmetry in the nebular brightening (as the nebula would only interact strongly with the ISM in the direction of its motion; \citealt{wareing07}).

\subsubsection{Nebular spectroscopy}
\label{subsubsec:neb_spec}

Longslit echelle spectroscopy was carried out using AAT-UCLES and NTT-EMMI, focusing on [O\,{\sc iii}] emission over eight different slit positions (see figure \ref{fig:lotr1} (b)) in order to gain velocity profiles across a good sample of the nebula. It is important to ascertain the true three-dimensional shape of the surrounding nebula in order to fully understand and constrain the shaping process, and subsequently give a clearer insight into mass-transfer mechanisms. Imaging is insufficient on its own due to a degeneracy between PN inclination and morphology -- for example, when the symmetry axis of a bipolar nebula is aligned perpendicular to the plane of the sky, it makes the nebula appear spherical \citep{kwok10}. If the nebular morphology is classified incorrectly then the fraction of aspherical PNe possessing a binary core will also end up being inaccurate, and this information is already exceedingly limited for longer period binaries due to selection biases for those stellar systems possessing wider orbits. Longslit spectroscopy can be used to acquire spatially resolved velocity maps of the constituent parts of the nebula in order to recover the `missing' third dimension of the morphology one cannot gain from imagery alone. The resulting spectra are then plotted as position-velocity (PV) arrays.

On 2005 March 03, a spectrum was acquired from the nebula using EMMI in its single order echelle mode, employing grating \#10 and a narrowband \OIII{} filter to prevent contamination from overlapping echelle orders. The maximum slit length of 330\arcsec{} and a slit width of 1\arcsec{} was used to give a resolution, R $\sim$ 54,000 (5.5\kms{}). A 1800s exposure was taken at a P.A. of 90$^{\circ}$ crossing the central star (slit position 1 in figure \ref{fig:lotr1}(b)), and the data were 2x2 binned to give a spatial scale of 0.33\arcsec{} and a velocity scale of 3.9\kms{} per pixel. The seeing was $\sim$0.7\arcsec{}.

On 2005 January 14, spectra were acquired from the nebula of LoTr~1 using the 79 lines/mm grating on the UCL Coud\'{e} Echelle Spectrograph (UCLES) of the Anglo-Australian Telescope (AAT). UCLES was operated in its longslit mode with a maximum slit length of 56\arcsec{} and a slit width of 1.97\arcsec{} to give a resolution R $\sim$ 20,000 (15\kms{}). The EEV2 CCD (2048$\times$4096 13.5 $\mu$m pixels) was used with binning of 2$\times$3, resulting in a pixel scale of 3.88\kms{} pixel$^{-1}$ in the spectral direction and 0.48\arcsec{} pixel$^{-1}$ in the spatial direction. 1800s exposures were taken at five different slit positions (shown in figure \ref{fig:lotr1}(b)) using a narrowband filter to isolate the 45$^{th}$ echelle order containing [O~\textsc{iii}] emission line profile. Slits 2--4 were taken at a position angle (P.A.) of 0$^{\circ}$, and slit 5 was taken at a P.A. of 90$^{\circ}$. The seeing during the observations was $\sim$2\arcsec{}. A further three slits (slit positions 6--8 in figure \ref{fig:lotr1}(b))were acquired using the same instrument and CCD on 2013 January 3 with a seeing of $\sim$1.5\arcsec{} and a binning of 2$\times$2 ($\equiv$ 0.32\arcsec{} pixel$^{-1}$ in the spatial direction). Here, a slit width of 1\arcsec{} was employed for slits 6 and 8 (R $\sim$ 45,000 $\equiv$ 6.7\kms{}) and 1.5\arcsec{} for slit 7 (R $\sim$ 30,000 $\equiv$ 10\kms{})\footnote{The slit width for slit 7 was altered to try to match the seeing conditions during the night. However, as all observations were carried out with the same binning in the spectral direction and the same grating (i.e. approximately the same dispersion), they can all be used to qualitatively assess the spatio-kinematic structure of the nebula (varying on scales $\gg$ the slit-widths employed, due to the angular size of the nebula also being $\gg$ the slit-width).}.

All the spectra were cleaned of cosmic rays and debiased appropriately. The UCLES spectra were wavelength calibrated against a ThAr emission-lamp, rescaled to a linear velocity scale appropriate for the \OIII{} emission, and corrected to heliocentric velocity, \vhel{}. Due to the optical set-up of EMMI, it was necessary to perform the wavelength calibrations using a long-exposure (3600s) ThAr emission lamp at the start and end of the night to gain a good number of arc lines, before cross-correlating with shorter-exposure (200s) Ne lamps taken immediately after each observation, to account for any drift due to telescope and instrument flexure (with small shifts accounted for with a linear correction). 

The reduced nebular spectra of LoTr~1 are presented in figures \ref{fig:lotr1_full_slits} and \ref{fig:lotr1_aat_1}, as position-velocity (PV) arrays. In each PV array, cross-section 0\arcsec{} defines where the central star is found. In figure \ref{fig:lotr1_full_slits} (a), the bright lines located around the central star at cross-sections +3\arcsec{} and $-$3\arcsec{} are most likely artefacts due to the comparative difference in brightness between the central star and the nebula \citep{jones10}.

The closed velocity ellipses shown in the PV arrays presented in figures \ref{fig:lotr1_full_slits} (a), \ref{fig:lotr1_aat_1} (c), and the central PV array of \ref{fig:lotr1_full_slits} (b) (i.e. representing slit positions 1, 4 and 7), have a major axis which has the same length as the diameter of the inner shell (see section \ref{subsubsec:lotr1_image}). This indicates that the nebular structure is indeed an isolated, closed shell rather than a projection effect related to a bipolar structure being viewed end-on. No significant asymmetries are observed in these velocity ellipses, which are consistent with a spherical shell or an elongated ovoid viewed directly along the symmetry axis. The expansion velocity, \vexp{} for the inner shell is measured to be 17$\pm$4\kms{} at the location of the central star, while the \vexp{} for the outer shell is measured to be 25$\pm$4\kms{} falling within the typical range for a PN (\citealt{weinberger89}). Assuming typical expansion properties for the nebula (i.e.\ velocity proportional to distance from the central star), the latter velocity is then the maximum expansion velocity (for an elliptical shell viewed pole-on) or the uniform expansion velocity of the shell (in the case of a sphere). The heliocentric systemic velocity, \vsys{} of this central shell was determined to be 14$\pm$4\kms{}.

In slit 1 (figure \ref{fig:lotr1_full_slits}(a)), emission is clearly detected from outside of the central shell at cross-sections $\sim$40\arcsec\ and $\sim-$40\arcsec{}, associated with the outer shell (see section \ref{subsubsec:lotr1_image}). Here, the emission from the eastern side appears blue-shifted with respect to the nebular \vsys{}, and the west appears red-shifted; this is indicative of an inclined and extended structure, e.g. an elliptical nebula, where the approaching ``lobe'' is tilted slightly to the east of its receding counterpart. Consideration of the other slits presented in \ref{fig:lotr1_aat_1} (b), (c) and (d), confirms this asymmetry in the velocity profile across the nebula, but indicates that the symmetry axis may lie closer to the northeast-southwest direction than east-west. However, any deviation from the line of sight must be rather small given the almost circular appearance of the shell in the images (figure \ref{fig:lotr1}).  

Determination of the exact structure and inclination of the nebula would require a more extensive, higher signal-to-noise (given the outer shells faintness) dataset, covering more of the physical extent of nebula, and a detailed spatio-kinematical model such as that presented in \citealp{jones12}. However, it is clear from both the imaging and spectroscopy presented here that LoTr~1 shows a double-shelled structure with evidence for an elliptical and slightly inclined outer shell, and a morphologically similar inner shell but with a different orientation.

\begin{figure}
\centering
\includegraphics[scale=0.60]{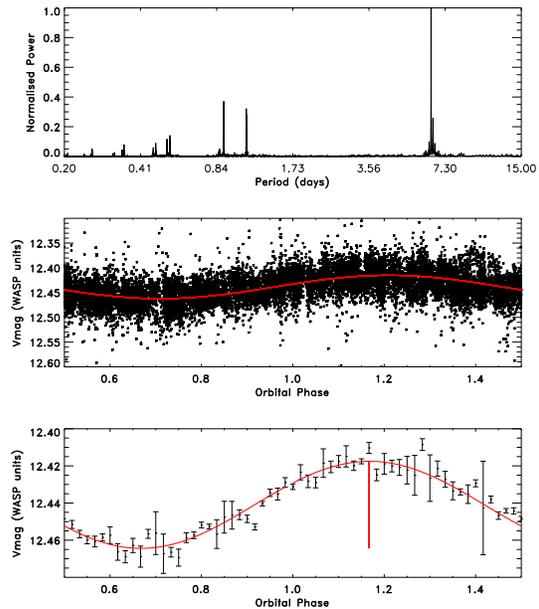}
\caption{SuperWASP photometry of LoTr~1.  Upper panel: Periodogram showing a clear detection of periodicity at $6.3967\pm0.0005$ days.  Middle panel: The data points folded on the period obtained from the periodogram. Lower panel: The data binned into phase bins, overlaid is a sinusoidal fit with a peak-to-peak amplitude of 0.061 magnitudes.}
\label{fig:superwasp}
\end{figure}

\begin{figure*}
\centering
\includegraphics[scale=0.9]{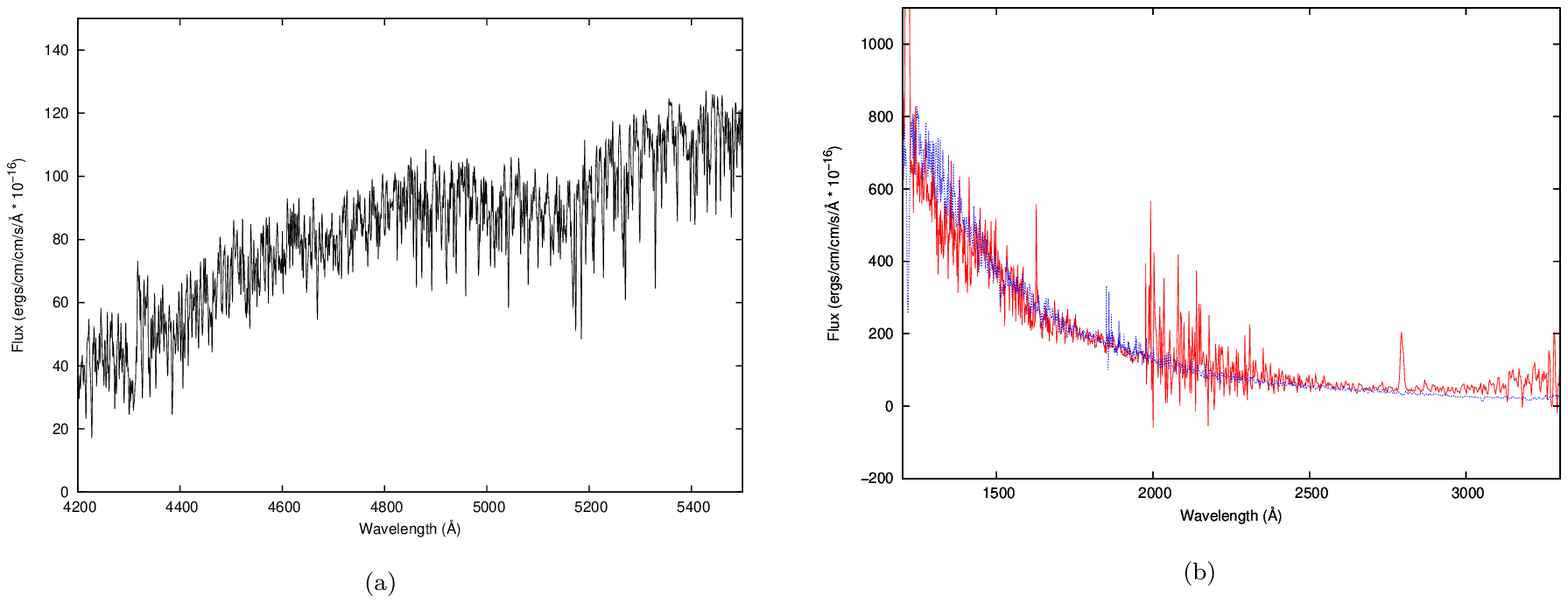}
\caption{(a) Flux-calibrated FORS2 spectrum of the central star system of LoTr~1. (b) Flux-calibrated IUE spectra of LoTr~1 (solid line), and of NGC~7293 (dashed line) used to determine the parameters of the white dwarf. Note also the presence of the Mg~II emission at 2800 \AA{} which is a sign of chromospheric activity from the cool companion (e.g. \citealp{jasn96,montez10})}
\label{fig:lotr1_stellar_2}
\end{figure*}

\subsubsection{Stellar photometry}
The field of LoTr~1 has been observed by WASP-S between 2006 May 4 and 2012 February 17 with a total of 21407 photometric points obtained with the two cameras DAS 226 and 228. WASP-S is a wide field survey camera situated at SAAO, Sutherland, South Africa, and together with its northerly sister, SuperWASP-N, is designed to obtain extremely accurate photometry of bright stars in order to search for transits from exoplanets (see \citealt{pollacco06}, for a more detailed description of the facility, the data reduction and archive). Data mining tools and the public archive are discussed by \citet{butters10}. 
 After reduction through the instrument pipeline \citep{pollacco06}, the time series were examined with a simple Lomb-Scargle filter for periodic signal detection. From the light curves reproduced in figure \ref{fig:superwasp} taken from \citet{jones11b}, a period of 6.4 days was derived with a peak-to-peak amplitude of 0.061 magnitude. This is close to the value of 6.6 days with an amplitude of 0.1 magnitude in V, initially derived by \citet{bond89}. Given the period, this amplitude is probably too high to be considered due to irradiation effects \citep{demarco06}, so the most likely explanation is that the periodicity is following spots on a cool star -- the signature of a rapid rotator, analogous to the systems of LoTr~5 (5.9$\pm$0.3d, \citealp{thevenin97}) and WeBo~1 (4.69$\pm$0.05, \citealp{bond03}).


\begin{figure}
\centering
\includegraphics[scale=0.35]{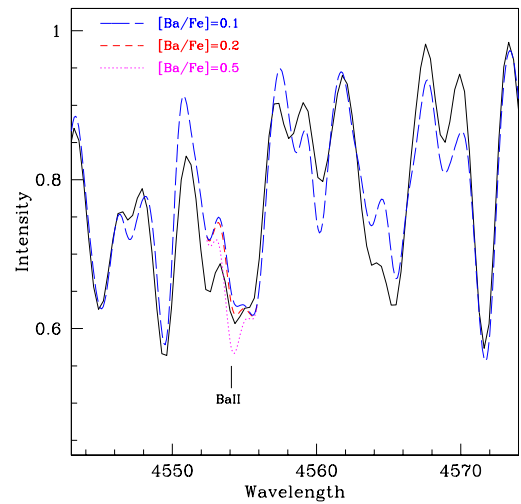}
\caption{LoTr~1 FORS2 stellar spectrum (solid line) alongside three synthetic spectra with [Ba/Fe] = 0.1, [Ba/Fe] = 0.2, and [Ba/Fe] = 0.5, smoothed to match the FORS2 resolution and plotted over a wavelength range which includes the \Ba{} 4554\AA{} line.}
\label{fig:Ba_wavelength}
\end{figure}

\begin{figure}
\centering
\includegraphics[scale=0.35]{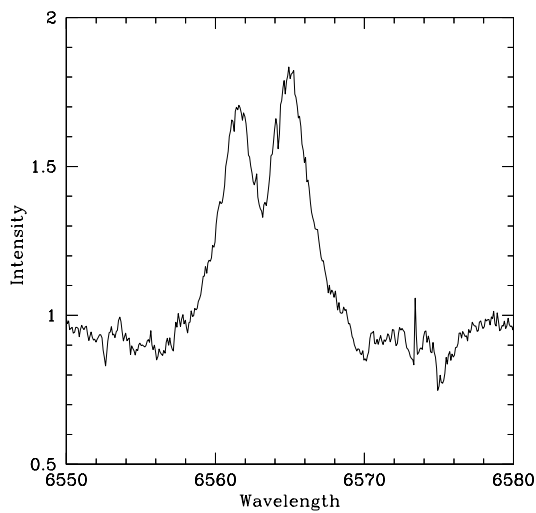}
\caption{H$\alpha$ emission seen in LoTr1.}
\label{fig:LotrHa}
\end{figure}

\subsubsection{Stellar spectroscopy}
 \label{subsubsec:lotr1_stel}

\begin{figure*}
\centering
\includegraphics{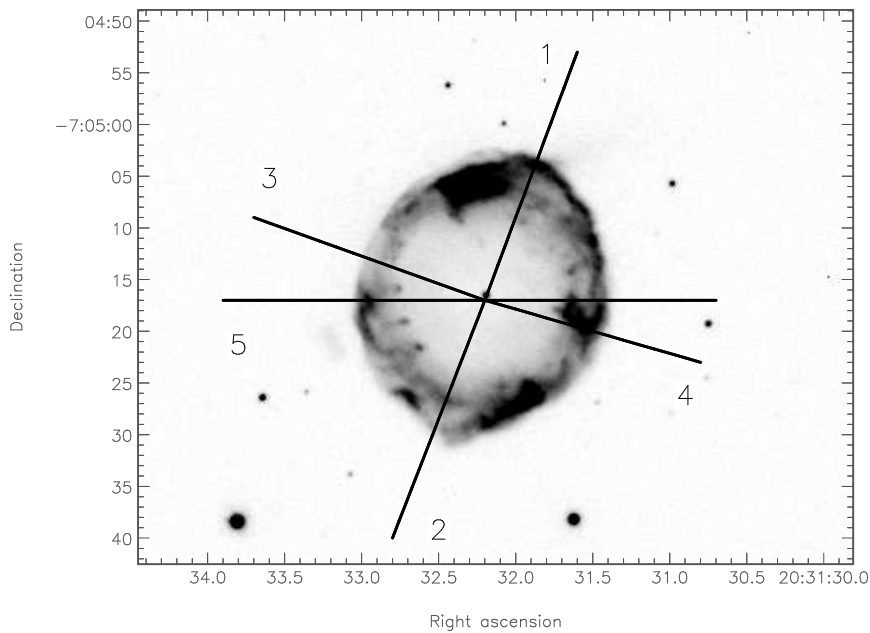}
\caption{\ha{}+\NII{} image of PN A70 showing all slit positions. Slits 1--4 were acquired in \OIII{} using VLT-UVES, and slit 5 was acquired in \ha{} and \OIII{} using SPM-MES. The central star is visible at $\alpha$ = 20:31:32.2, $\delta = -$07:05:17.0.}
\label{fig:a70}
\end{figure*}

On 2012 February 10, a low-resolution spectrum was acquired of the central star of LoTr~1 using the FOcal Reducer and low dispersion Spectrograph (FORS2; \citealt{appenzeller98}) with grism 1200g, on the Antu unit (UT1) of the Very Large Telescope (VLT) based at ESO-Paranal. FORS2 was operated with a maximum slit length of 6\arcmin{} and a 0.5\arcsec{} slit width (R $\sim$ 3000), and at a position angle of 45\degr{}. The data were 2$\times$2 binned to give a spatial scale of 0.25\arcsec{} pixel$^{-1}$, with a dispersion on the spectral axis of 0.3\AA{} pixel$^{-1}$. A single 600s exposure was taken under seeing of $\sim$ 0.7\arcsec{}.

The stellar data were extracted to 1-D and flux-calibrated against standard star Hz~4, and are presented in figure \ref{fig:lotr1_stellar_2} (a). All reductions were carried out using standard \textsc{starlink} routines.
 
 
The central star of LoTr~1 was classified as a K1~III-type giant by comparing the flux-calibrated FORS2 spectrum with UVES POP standard stars \citep{bagnulo03} that had been rebinned and smoothed to match the resolution of the FORS2 data. To refine the analysis and derive the possible s-process overabundance, we used the stellar spectral synthesis code of R. Gray, \textsc {spectrum} version 2.76\footnote{See http://www1.appstate.edu/dept/physics/spectrum/}, with models from \citet{castelli03}. The best fit parameters were determined to be \teff{} $= 4750 \pm$ 150  K and log $g = 2.0 \pm 0.5$, which is in agreement with the K1~III spectral type. Using the average value of the absolute magnitude, M$_{v}$ = +0.7 for a giant of this spectral type as given in Allen's `Astrophysical Quantities' (2000), the flux for the spectrum shown in figure \ref{fig:lotr1_stellar_2} (a) that allows the derivation of an apparent magnitude, m$_{v}$  = +12.6 (accounting for 20\% slit-losses), and extinction $A_V$=0.1285 \citep{schlegel98}, one can determine a rough distance to the star of 2.6 kpc and a not-unreasonable radius of the giant of 11.5 R$_\odot$. Alternatively, one can use the average value of $V$ derived from SuperWASP (see figure \ref{fig:superwasp}), i.e. $V=12.44$, to derive a distance of 2.1 kpc. Note that this observed SuperWASP magnitude is likely to be contaminated by line emission and close field stars because of the broad observing band and large (14\arcsec{}) pixels. Therefore, we think it is preferable to use the distance derived from the flux calibrated spectra.


To check that the giant and the planetary nebula are indeed linked and not merely a chance superposition, we computed the radial velocity difference between the nebular lines in the FORS2 spectrum and the giant star's absorption lines, cross-correlating our spectrum (shifted so that the nebular lines are at zero velocity) with our synthetic K1~III spectrum. The resulting stellar velocity of 4$\pm$2\kms{} with respect to the nebula implies that the cool central star and nebula are physically related.

\begin{figure*}
\centering
\includegraphics[scale=0.9]{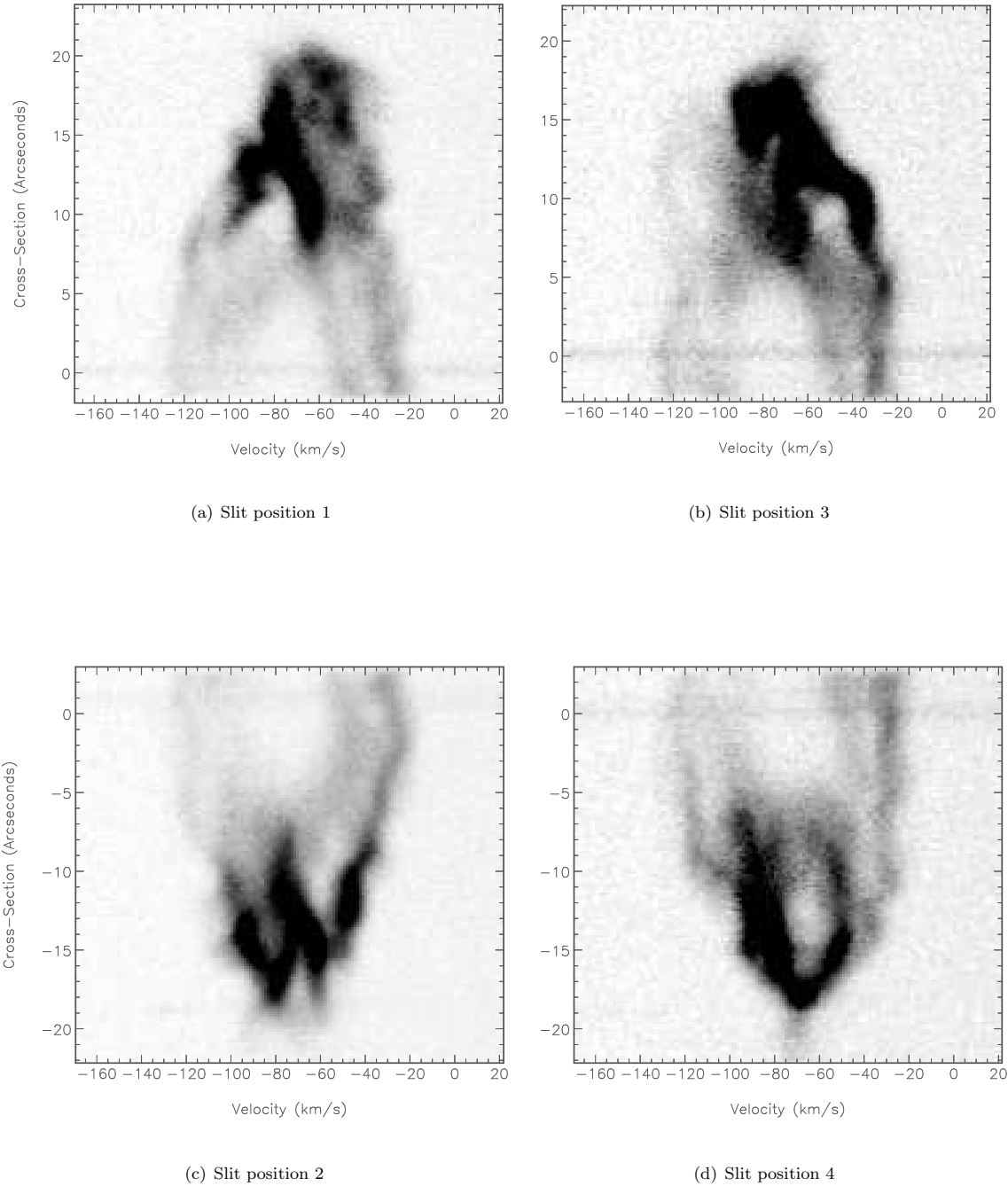}
\caption{PV arrays showing reduced VLT-UVES spectra in \OIII{} from A70. Figures (a) and (c) show emission from the major axis, figures (b) and (d) are from the minor axis. Positive spatial
offsets are to the northern (1 and 2) or eastern (3 and 4) ends of the
slits.The velocity axis is heliocentric velocity, \vhel{}. The display scale has been modified to highlight the spatio-kinematic features referred to in the text. Cross-section 0\arcsec{} defines where the central star is found.}
\label{fig:a70_spec}
\end{figure*}

The FORS2 spectrum was used to check for signs of \Ba{} pollution at 4554\AA{}. Figure \ref{fig:Ba_wavelength} shows the FORS2 spectrum of LoTr~1 plotted alongside three synthetic spectra with various barium enhancements, from [Ba/Fe] = 0.1 to [Ba/Fe] = 0.5. An over-abundance of 0.5 or greater classifies the system as a definite \Ba{} star, with a value of 0.2--0.5 being possibly a `mild' \Ba{} star \citep{pilachowski77}. Due to the relatively low S/N and resolution of the spectrum, we are unable to determine a definitive value for the barium abundance in LoTr~1, but we can clearly state that it is much less than 0.5, and therefore does not show any measurable barium enhancement. This is contrary to both A70 and WeBo~1 (see \citealt{miszalski12} and \citealt{bond02}, respectively), which both possess definite \Ba{} stars.

On 2013 January 3, 3$\times$30 minute spectra were acquired from the CSPN of LoTr~1 using AAT-UCLES with the 79 lines/mm grating, operated in full echelle mode with a slit width of 1\arcsec{} to give a resolution R $\sim$ 45,000.  All spectra were reduced using standard \textsc{starlink} routines, corrected to heliocentric velocity and then summed.  Unfortunately, because of the large interorders in the spectra\footnote{UCLES operated in this mode gave non-continuous wavelength coverage from roughly 5200\AA{} to 8900\AA{} across 19 orders with interorder spacing of 100--200\AA{}.}, it is not possible to assess the barium abundance, as none of the available orders contain the Ba\textsc{ii} lines at 6141.7\AA{} and 6496.9\AA{}. We have verified, however, that the iron lines are well fitted with a solar abundance and our preferred model, in agreement with what we derived from the FORS2 spectrum. Looking at the La \textsc{ii} 6390\AA{} and Y\textsc{ii} 6222\AA{} lines, it is also clear that these elements are not overabundant, confirming the lack of s-process enhancement in LoTr~1.

The AAT spectra allowed us to find that the stellar component of H$\alpha$ is in emission (see figure \ref{fig:LotrHa}). The line is very broad and is clearly double-peaked, with an equivalent width of about 6\AA{} and the velocity spread is 572 km/s. Using the method of \citet{hodgkin95}, we derive an H$\alpha$ luminosity $L_{\rm H\alpha}$ = 0.044 L$_\odot$. With the above estimated total luminosity of the star ($L_*\sim$ 60 $L_\odot$), this gives a value $\log L_{\rm H\alpha}/L_*= -3.12$. 

Such H$\alpha$ double-peaked emission lines have been found in the other stars we are concerned with here, LoTr~5 \citep{jasniewicz94,strassmeier97} and A35 \citep{acker90a,jasniewicz92}, but the origin is still unknown. Rapidly rotating giants, such as RS CVn or FK Com stars, are known to have high chromospheric activity which is often revealed by emission cores in some lines. And indeed, \citet{acker90a} find a modulation of the H$\alpha$ emission line with the rotation phase, while \citet{jasniewicz92} postulate that the variable double-peaked emission line is the result of an overlap between an absorption and an emission line at H$\alpha$, with the possibility for the absorption component to be formed in the photosphere or through a self-absorption process as in Be stars. However, the FWHM velocity we measure seems too high to be caused by mass motions inside the chromosphere, while the luminosity is too small to be due to accretion. 
Such double-peaked emission are also sometimes found in symbiotic stars -- detached systems which interact via wind accretion. A few of them, out of an outburst event, produce bipolar nebulae very similar to planetary nebulae. It turns out that many symbiotic stars show double H$\alpha$ profiles (e.g. \citealp{schild96, burmeister09}), which might be caused by a narrow absorption component from the giant overlaid with very broad H$alpha$ emission from high-velocity jets at the core of the system, or by disc-like structures. It would thus be of interest to further study the possible link between LoTr~1 and symbiotic stars.

A spectrum of the central star system of LoTr~1 was acquired using the International Ultraviolet Explorer (IUE) satellite by \citet{bond89}. Presented here in figure \ref{fig:lotr1_stellar_2} (b), it indicates a strong UV continuum of \teff{} $\ge$ 100 kK. Comparing this IUE spectrum to a known WD within PN NGC~7293 (d = 219$^{+27}_{-21}$~pc, m$_{v}$ = +13.5; \citealp{harris06}) gave us a reasonable fit: \teff{} $\ge$ 123 kK, and R = 0.017\rsun{}. Based on values of D = 2\arcmin{}22\arcsec{}$\pm$4\arcsec{}, and \vexp{} = 25$\pm$4\kms{} at a distance of 2.1--2.6~kpc, we derived a kinematical age of 33,000$\pm$9,000 years for the outer nebular shell. For the inner shell, given values of D = 47$\pm$2\arcsec{} and \vexp{} = 17$\pm$4\kms{},  the age is derived as 16,000$\pm$6,500 years. Using the white dwarf evolutionary curves from \citet{bloecker95} and assuming an average remnant mass of 0.6\msun{}, we derived a stellar temperature at the age of the PN of approximately 120,000 K. The derived radius and age of LoTr~1 are consistent with Bloecker's evolutionary curves.

\subsection{A70}
 \label{subsec:A70}

		\subsubsection{Imaging}
		 \label{subsec:a70_obs}

The \ha{}+\NII{} image of A70 shown in figure \ref{fig:a70} was acquired on 2012 July 08, using FORS2 under program ID 0.89.D-0453(A), with an exposure time of 60s and seeing of 0.8\arcsec{}. At first glance the image shows a general ring-like appearance similar to that of other ring-like PNe (e.g.\ SuWt~2: \citealt{jones10}); however, just as noted by \citet{miszalski12}, a closer inspection reveals a ``ridged'' profile more like that of a bipolar nebula viewed end-on (e.g. Sp~1: \citealt{jones11a}). Furthermore, this image shows in detail the low-ionisation knots first identified by \citet{miszalski12}.  Many of these structures seem to be akin to the cometary globules seen in the Helix Nebula (dense condensations of molecular gas embedded in the ionised nebula, see \citealt{meaburn92}), with knotty heads closest to the nebula centre and extended tails reaching out towards the outer rim. Extended material is also visible outside the east-southeasterly edge of the nebular ring (the emission visible to the north of the ring originates from a background field galaxy).

Assuming the ``ring'' is a physical structure rather than a projection effect, one can deduce the inclination of the nebula by deprojection. The angular size of the nebula was determined to be 44\arcsec $\times$ 38\arcsec{}$\pm$2\arcsec{}, falling in line with the previously given value of 45.2\arcsec x 37.8\arcsec{} by \citet{tylenda03}, giving an inclination of $30$\degr{}$\pm10$\degr{}.

\begin{figure}
\centering
\includegraphics[scale=0.8]{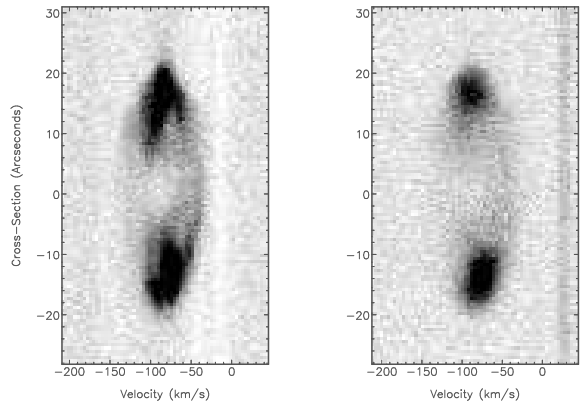}
\caption{PV arrays showing reduced SPM-MES spectra of A70. Emission is through the horizontal axis (slit 5 in figure \ref{fig:a70}). The velocity axis is heliocentric velocity, \vhel{}. The display scale has been modified to highlight the spatio-kinematic features referred to in the text. Cross-section 0\arcsec{} defines where the central star is found.}
\label{fig:a70_spec_spm}
\end{figure}


			\subsubsection{Nebular Spectroscopy}
			 \label{subsubsec:a70_neb}

On 2011 June 10-11, high resolution data of the nebula of A70 were acquired in [O\,{\sc iii}] using grating \#3 on the visual-to-red arm of the Ultraviolet and Visual Echelle Spectrograph (UVES) on Kueyen Unit (UT2) of the VLT \citep{dekker00}, under program ID 087.D-0174(A). UVES was operated in its 30\arcsec{} longslit mode with a 0.6\arcsec{} slitwidth (R $\sim$ 70,000, 4.3\kms{} pixel$^{-1}$) to give a spatial scale of 0.17\arcsec{} pixel$^{-1}$. A filter was used to isolate the [O\,{\sc iii}] emission lines and prevent contamination from overlapping orders. The seeing was between 0.5\arcsec{} and 0.7\arcsec{} for all observations. Four 1200s exposures were taken over four different slit positions. Slits 1 and 2 were taken with VLT-UVES at a P.A. of 160${^\circ}$ and slits 3 and 4 at a P.A. of 70${^\circ}$, to line up with the major and minor axis of the nebula, respectively. Slit 5 was acquired in both [O\,{\sc iii}] and \ha{} on 2011 May 15 using the Manchester Echelle Spectrograph (MES) mounted on the 2.1-m San Pedro Martir (SPM) telescope based at the Observatorio Astronomico Nacional in Mexico \citep{meaburn03, lopez12}. The full slit length of 5\arcmin{} was used with a slitwidth of 150$\mu$m ($\equiv$ 2\arcsec{}, R $\sim$ 30,000), and taken at a P.A. of 90${^\circ}$. The data were 2x2 binned to give a spatial scale of 0.75\arcsec{} pixel$^{-1}$. The seeing was $\sim$ 3\arcsec{}

The nebular spectra acquired from VLT-UVES shown in figure \ref{fig:a70_spec} show two highly filamentary components, one red-shifted and one blue-shifted, joined by bright knots of emission where the slits cross the nebular ring, to form a closed velocity ellipse in both axes. These filamentary and irregular structures are typical of disrupted nebulae, where instabilities have begun to structurally deform the shell \citep{guerrero12}, and clearly show that A70 is not simply an inclined ring but instead has ``bubbles'' extending in the line of sight. It is reasonable to assume that these bubbles form a closed (as the velocity ellipses are closed along both axes) and axisymmetric (the blue- and red-shifted components are roughly symmetrical) structure. The bright emission at the extremes of each slit indicate that the nebula may have a cusped waist, with slits 1 and 2 showing the ``crow's foot''-like structure typical of narrow waisted nebulae viewed along their symmetry axis \citep{jones12}. However, there is a clear brightening in these regions and those of slits 3 and 4, consistent with a bright ring. We therefore deduce that A70 comprises of such a bright ring, encircling the waist of a disrupted and faint bipolar shell.

Using the same spectra, a polar expansion velocity \vexp{} for A70 was calculated to be 39$\pm$10\kms{}. This is in agreement with the value for expansion velocity of \vexp{} = 38\kms{} given by \citet{meatheringham88}, although no error was quoted. A \vsys{} of $-$73$\pm$4\kms{} was determined for the nebula,  which is consistent with the value of $-$72$\pm$3\kms{} by \citet{miszalski12}. The kinematical age of A70 was determined to be 2700$\pm$950 years kpc$^{-1}$. Taking the distance to the nebula to be 5 kpc, as given by \citet{miszalski12}, this gives an overall kinematical age for A70 of order 13,400$\pm$4,700 years.

The limited depth and resolution of the MES-SPM spectra offer little extra to the discussion of the nebular morphology and kinematics, other than to show that there is no major difference between the  emission in \ha\ and that in \OIII{}. They are shown here in figure \ref{fig:a70_spec_spm} mainly for comparison with similar data acquired from PN WeBo~1, which will be discussed in section \ref{subsec:webo1}.

			 

\subsection{WeBo~1}
 \label{subsec:webo1}

\begin{figure*}
\centering
\hspace*{-0.3cm}
\includegraphics[scale=0.9]{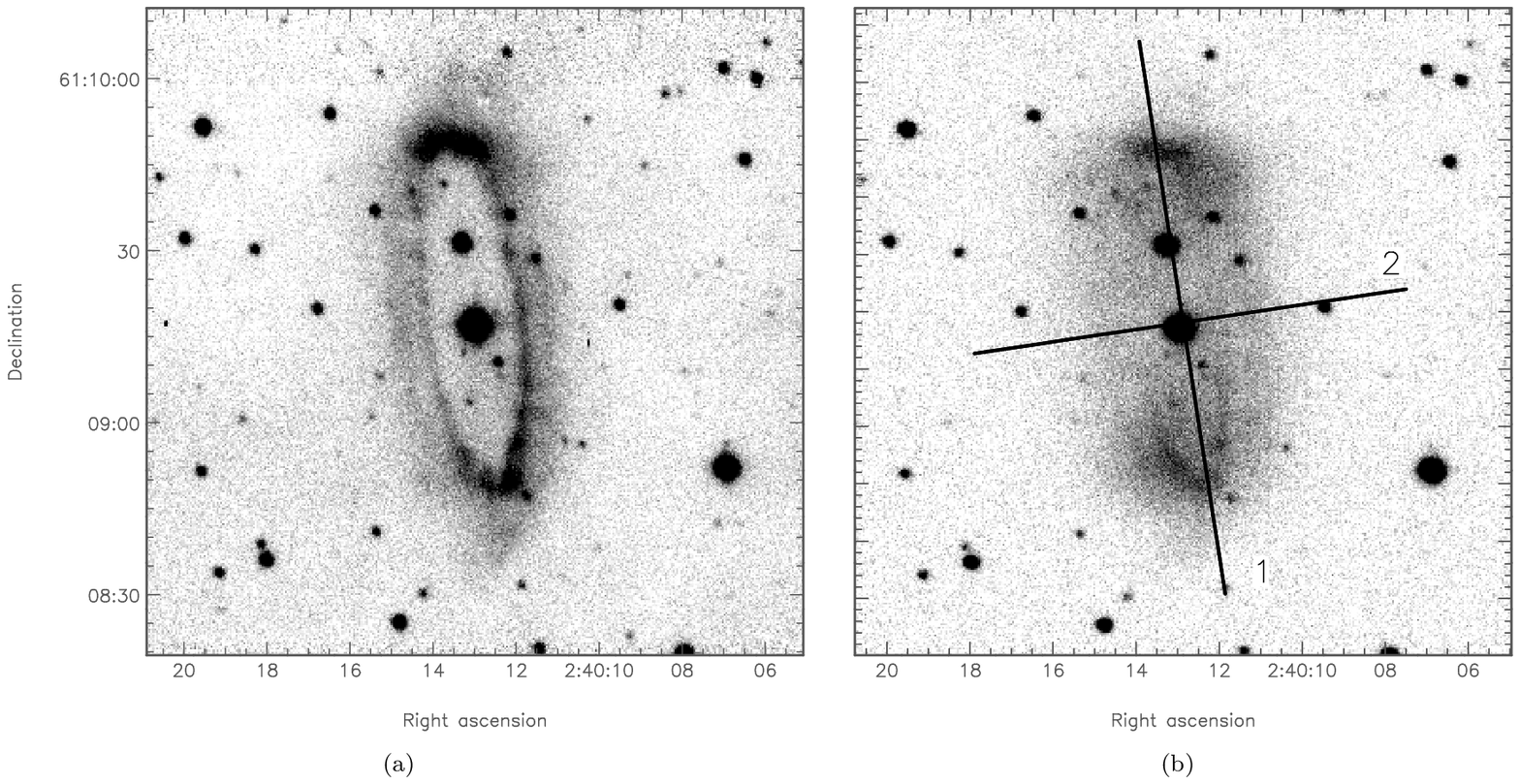}
\caption{Images of PN WeBo~1 in (a) \ha{}+\NII{}, and (b) \OIII{}, with (b) showing the two slit positions. Both slits were acquired in [N\,{\sc ii}] and [O\,{\sc iii}] using SPM-MES. The central star is visible at $\alpha$ = 02:40:13.0, $\delta$ = +61:09:17.0. North is to the top of the images, East is left.}
\label{fig:webo1}
\end{figure*}

		\subsubsection{Imaging}
		 \label{subsubsec:webo1_obs}

The deep \ha{}+\NII{} image shown in figure \ref{fig:webo1} (a) is the result of coadding 2 $\times$ 120s exposures, each with seeing better than 1.2\arcsec{}, acquired as part of the IPHAS survey \citep{drew05} using the Wide Field Camera (WFC) on the 2.5m Isaac Newton Telescope based at the Observatorio Roque de los Muchachos, La Palma.  The \OIII{} image shown in figure \ref{fig:webo1} (b) was acquired using the same instrument on 2010 September 9, with an exposure time of 1200s and under seeing of 1.4\arcsec{}.

The images show another ring-like morphology, although structurally different to A70 (see section \ref{subsec:A70}), with a pronounced inner edge and fainter, more extended emission around its entire circumference -- the ring is particularly diffuse in \OIII{}, as shown by the lack of a visible inner edge. Similar extended emission is also found in SuWt~2 \citep{jones10} and HaTr~10 \citep{tajitsu99}, where the ring is actually the waist of an extended bipolar structure, possibly indicating that WeBo~1 may display the same morphology but with as yet undetected, very faint lobes. In SuWt~2, \citet{jones10} attribute this extended material to structural and brightness variations across an irregular toroidal structure; however (particularly in the light of the {H$\alpha$}+[N\,{\sc ii}] spectra acquired - see figure \ref{fig:webo1_spec}), WeBo~1 shows a much more regular and even ring-like shape, indicating that this is more likely an intrinsic structural property (i.e. a tear-drop rather than circular shaped cross-section).

\citet{smith07} deprojected the ring of WeBo~1, determining that it is seen almost edge-on with an inclination of 75\degr$\pm$3\degr{} with an inner-ring radius of $\sim$ 25\arcsec{}, which is consistent with the dimensions of the ring as cited by \citealt{bond03} (64\arcsec x 22\arcsec{}), and as measured from the images presented here (65\arcsec{}$\times$20\arcsec{}$\pm$4\arcsec{}).

			\subsubsection{Nebular Spectroscopy}
			 \label{subsubsec:webo1_neb}

\begin{figure*}
\centering
\includegraphics{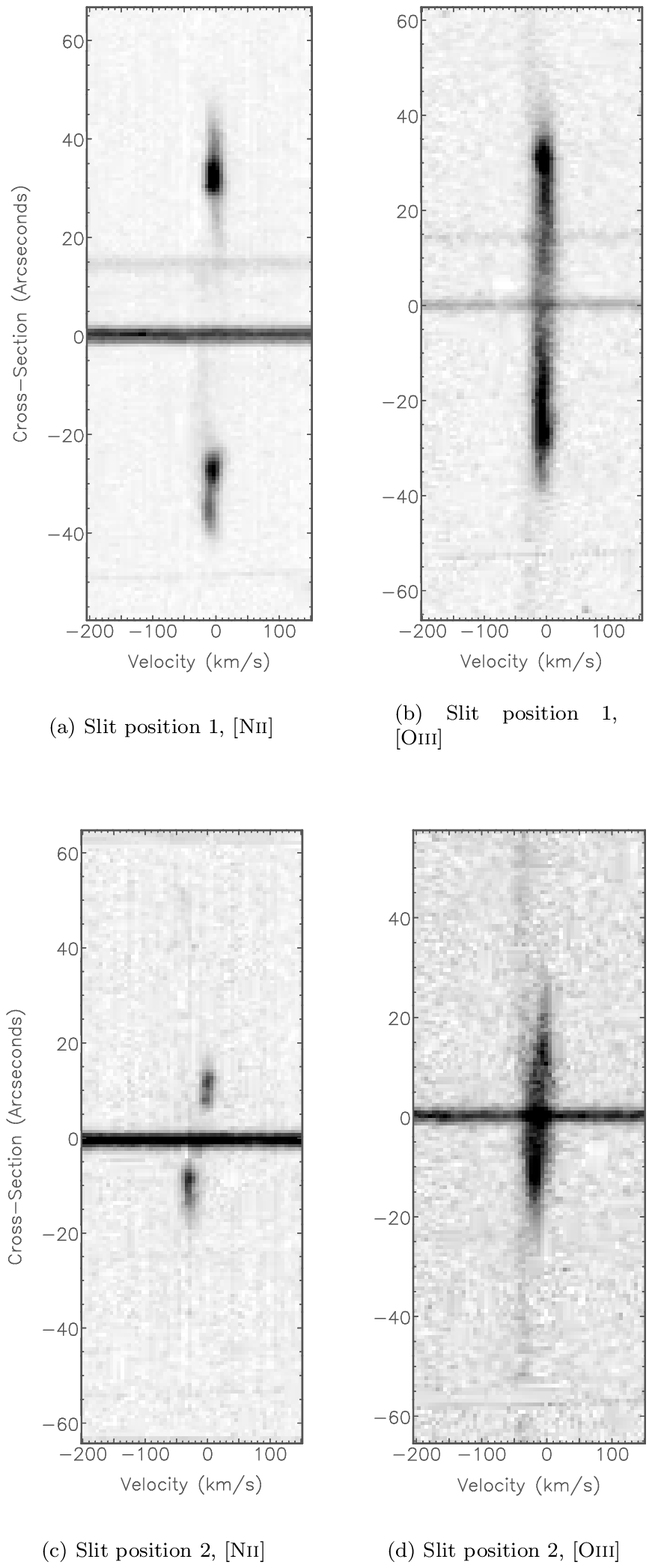}
\caption{PV arrays showing reduced \NII{} and \OIII{} SPM-MES spectra of WeBo~1. Figures (a) and (b) show the [N\,{\sc ii}] and [O\,{\sc iii}] emission from Slit 1, (c) and (d) from Slit 2 (see figure \ref{fig:webo1}). North is to the top of the array. The velocity axis is heliocentric velocity, \vhel{}. The display scale has been modified to highlight the spatio-kinematic features referred to in the text. Cross-section 0\arcsec{} defines where the central star is found. The continuum of a field star is visible at cross-section +15\arcsec{} in figures (a) and (b).}
\label{fig:webo1_spec}
\end{figure*}

On 2010 December 10, spectra were acquired of WeBo~1 in both {H$\alpha$}+[N\,{\sc ii}] and \OIII{} using SPM-MES. The maximum slit length of 5\arcmin{} was used with a slitwidth of 150$\mu$m ($\equiv$ 2\arcsec{}, R $\sim$ 30000) for each filter. The data were 2x2 binned to give a spatial scale of 0.75\arcsec{} pixel$^{-1}$. Slit 1 was taken at a P.A. of 353${^\circ}$ and slit 2 was taken at a P.A. of 263${^\circ}$ to cover the major and minor axes of the nebula, respectively. The seeing was $\sim$1.5\arcsec{}. Due to the \ha{} profiles having high galactic background emission, only the background-subtracted [N\,{\sc ii}] and \OIII{} are presented here. The reduced PV arrays are presented in figure \ref{fig:webo1_spec}.

Just as in the imagery (see section \ref{subsubsec:webo1_obs}), the \NII{} profiles show well defined emission originating from the ring, while the \OIII{} is much more diffuse. The \NII{} profiles from both slit positions show strong emission at the inner edge of the ring, with slightly fainter material (also with a lower velocity dispersion) reaching out to greater angular extents -- consistent with the ``tear-drop'' cross-section interpretation presented in section \ref{subsubsec:webo1_obs}. The \NII{} profile from slit 1 shows little velocity difference between the emission originating from the two opposing sides of the ring, indicating that the slit is roughly perpendicular to the P.A. of the nebular symmetry axis; therefore, the deprojected velocity difference of the two sides of the ring in slit 2 should offer a good measure of the ring's expansion velocity. Taking the nebula inclination to be 75\degr{}, the deprojected expansion velocity for the ring was calculated to be \vexp{} = 22.6$\pm$10\kms{}.



A \vsys{} of -6$\pm$4\kms{} was calculated for the nebula. The kinematical age of WeBo~1 was determined to be of order 7300$\pm$3700 years kpc$^{-1}$. Taking the distance to the nebula to be 1.6 kpc \citep{bond03}, this gives an overall age for WeBo~1 of 11,700$\pm$5,900 years.\footnote{\citet{bond03} assume an expansion velocity of 20\kms{} to derive a similar age of 12,000$\pm$6,000 yrs.}

No clear evidence of lobes or extended nebular structure are detected in the spectra, unlike for A70 (see section \ref{subsubsec:a70_neb}); however, faint material detected inside the ring on the \NII{} PV array of slit 1 could be consistent with such a structure.  Deeper, higher resolution spectra are required to confirm the nature of this emission, but it is safe to say that if any lobes are present they are significantly fainter than those of A70, as its lobes were still clearly detected by SPM-MES spectra (see figure \ref{fig:a70_spec_spm}).

\begin{table*}
\caption{Physical parameters of LoTr~1, A~70 and WeBo~1}
\label{tab:params}
\begin{tabular}{lcccl}
\hline
Nebula & $v_{exp}$ (\kms{}) & Kinematical age (yrs) & Physical size (pc) & Morphology\\
\hline
\vspace*{0.5cm}LoTr~1 (inner)& 17$\pm$4 & 17,000$\pm$5,500 & 0.59$\pm$0.05 & spherical/elliptical\\
\vspace*{0.5cm}LoTr~1 (outer)& 25$\pm$4 & 35,000$\pm$7,000 & 1.86$^{+0.05}_{-0.09}$ & spherical/elliptical\\
\vspace*{0.5cm}A~70 & 39$\pm$10 & 13,400$\pm$4,700 & 1.10$^{+0.06}_{-0.04}$ & Ringed-waist with detected lobes\\
\vspace*{0.2cm}WeBo~1 & 23$\pm$10 & 11,700$\pm$5,900 & 0.50$\pm$0.03 & Ringed-waist, no lobes detected\\
\hline
\end{tabular}
\end{table*}

\section{Discussion}

\subsection{The Abell-35 group and PN mimics}
\label{subsec:mimics}

Abell 35 - the archetype of the class of PNe discussed in this paper -- has recently been shown to be a PN mimic.  It is, therefore, critical to establish the true nature of the objects considered here before beginning a comparison. We restrict our analysis to the three PNe presented in this work, and exclude both Abell~35 (which is no longer considered a true PN) and LoTr~5 (which is a considerably more complex case and discussed in more detail in \citealp{frew08}; see also \citealp{graham04}). 

 \citet{frew10} present a `recipe' for determining whether we can classify an object as a true PN or not, parts of which we can apply to the nebulae presented here:

\begin{itemize}
\item Presence of a hot, blue central star: All three nebulae presented here show evidence of excess UV flux that point towards the existence of a hot companion -- see Fig.~5a for LoTr~1, \citealt{miszalski12} for A70, and \citealt{siegel12} for WeBo~1.
\item Nebular morphology: Each PN possesses what we would classify as a `typical' PN shape, with rings (A70 and WeBo~1) and shells (LoTr~1). A mimic is often more diffuse.
\item Systemic velocity: The \vsys{} of the central star is consistent with the \vsys{} of the nebula for both LoTr~1 and A70 (see section \ref{subsubsec:lotr1_stel}, and \citealp{miszalski12}), and so we can say that the observed emission comes from a true planetary nebula. No \vsys{} for the central star of WeBo~1 has been published.
\item Nebula expansion: All three nebulae have been shown to have an expansion velocity typical for a PN (see Table \ref{tab:params}).
\item Nebular diameter: Using the values stated in this paper, the physical diameters of LoTr~1, A70 and WeBo~1 are all of order $\sim$1 pc (see Table \ref{tab:params}): a sensible value for a PN.
\item Galactic latitude: Two of the three nebulae are found at high galactic latitudes of -22\degr{} and -25\degr{} for LoTr~1 and A70, respectively (WeBo~1 is at and +1\degr{}). PNe are more likely to be found away from the Galactic plane than isolated Str\"omgren spheres.
\end{itemize}

We can thus be confident that the three objects studied in this paper show the characteristics of bona-fide planetary nebulae, although, as mentioned earlier, their link with the class of symbiotic stars should also be investigated further. 

\subsection{Conclusions}
\label{subsec:conclusions}

From the study conducted in this paper, we have been able to show that LoTr~1 possesses a double-shelled, slightly elliptical morphology of age of 35,000$\pm$7,000 years for the outer shell, and 17,000$\pm$5,500 years for the inner. We have been able to infer the presence of a K1~III-type giant (\teff{} $\sim$ 4500~K) and hot white dwarf (\teff{} $\sim$ 123~kK, R = 0.017\rsun{}) binary system at its core. The cool star has been shown to be kinematically associated with the nebula, and to have a rotation period of 6.4 days. Although it was not possible to accurately determine the [Ba/Fe] value for the central star system, we were able to say with confidence that LoTr~1 does \emph{not} show any evidence for an over-abundance of \Ba{}. LoTr~1 also presents double-peaked emission lines, which have been seen in the other PNe with cool central stars.

Unlike LoTr~1, the PNe A70 and WeBo~1 have both been previously confirmed to contain a \Ba{}-enriched central star system at their core. The two nebulae are also shown here to display morphologies distinct to that of LoTr~1, with both possessing ring-like waists and possible extended lobes. The similar morphologies and chemical enrichment strongly imply that the two have undergone very similar evolutionary or mass-loss processes. It is possible that the wind-accretion process involved in the formation of \Ba{} stars is also responsible for the formation of these ring-like morphologies. Although the CSPN of LoTr~1 does share some common traits with those of A70 and WeBo~1 -- namely binarity with a hot- and cool-components, and rapid rotation of the secondary -- both the lack of a significant over-abundance of \Ba{} and the marked difference in nebular morphology would imply a difference in the evolution of this system. The lack of \Ba{} enhancement could be explained by a difference in progenitor mass, metallicity, or simply quantity of mass transferred via the same wind-accretion process (the amount of material accreted is strongly dependent on orbital separation -- \citealp{boffin88}). However, as shown by \citet{boffin94} only a small amount of matter is needed to be accreted to make a star appear as a barium star and some mass must have been transferred as it is required in order to spin up the secondary to its rapid rotator state. The most obvious explanation, therefore, is that the mass was transferred at an earlier stage in the evolution of the primary, i.e. before the thermally-pulsing AGB phase, when the s-process elements are created and brought to the surface. This would allow us to infer that the AGB evolution of the primary was cut short by this mass-transfer episode, signs of which should be detectable in the properties of the WD. We strongly encourage follow-up observations of the system in order to confirm this hypothesis, and in particular it would be crucial to determine the orbital period of these systems. It is, however, important to note that given the inclination of the LoTr~1 nebula (very close to pole-on), any radial velocity variations of the central star system would be very difficult to detect, particularly for the expected period of $\sim$1--3 years, and this therefore requires very high spectral resolution and stable instruments. 

Coming back to the nebula, multiple shells are not uncommon in planetary nebulae, with 25-60\% found to show outer structures \citep{chu87}. However, it is important to distinguish here between `halo-like' shells \citep{corradi03,corradi04}, which are extended, generally spherical, structures attached to the inner shell, and detached outer shells, with the latter being far less common. This is critical as the haloes are generally understood to be the ionised remnant of mass lost on the AGB which is now being swept up to form the inner shell, while the formation mechanism for multiple, detached shells is still a mystery. \citet{schoenberner97} show that this may be possible via a combination of photo-ionisation and wind interaction, or, alternatively, a binary evolution might be responsible for rapid changes in mass-loss that could form two distinct shells, such as with Abell~65 \citep{huckvale13}.

LoTr~1 clearly shows a detached outer shell which may have been produced by rapid changes in mass-loss/transfer in the CSPN. The difference in kinematical ages between the two shells, of roughly 18,000-20,000 years, is much shorter than the single star evolutionary timescales on which these changes might occur. 

Perhaps the nature of LoTr~5 should also be investigated further, as it too does not have an apparent ring-like morphology despite possessing a rapidly rotating, G5~III-type, \Ba{}-rich central star. As mentioned earlier, \citet{montez10} have carried out a study into the x-ray emission emanating from this system and concluded that it is most likely chromospheric in origin, implying the presence of a spun-up companion.

The results presented here show that we still have some way to go to fully constrain mass-transfer mechanisms in intermediate-period binary and post-AGB systems, and further study of other such similar systems would be highly beneficial, in particular with regards to \Ba{} pollution.


\section*{Acknowledgments}
AAT gratefully acknowledges the support of STFC and ESO through her respective studentships. This work was co-funded under the Marie Curie Actions of the European Commission (FP7-COFUND). Based on observations made with ESO telescopes at the La Silla Paranal Observatory under programme IDs: 0.74.D-0373(A), 087.D-0174(A), 087.D-0205(A), 088.D-0750(A), 0.88D-0573(A), and 0.89D-0453(A). We thank the staff at the ESO La Silla Paranal Observatory and the Anglo-Australian telescope for their support in the acquisition of the observations. WeBo~1 images were acquired with the 2.5m Isaac Newton Telescope located in the Spanish Observatorio del Roque de los Muchachos on La Palma, Canary Islands, which is operated by the Instituto de Astrof\'sica de Canarias (IAC). Both A70 and WeBo~1 analyses are based on observations obtained at the Observatorio Astron\'omico Nacional at San Pedro Martir, Baja California, Mexico, operated by the Instituto de Astronom\'a, Universidad Nacional Aut\'onoma de M\'exico.

Some of the data presented in this paper were obtained from the Mikulski Archive for Space Telescopes (MAST). STScI is operated by the Association of Universities for Research in Astronomy, Inc., under NASA contract NAS5-26555. Support for MAST for non-HST data is provided by the NASA Office of Space Science via grant NNX09AF08G and by other grants and contracts.

\bibliographystyle{mn2e}
\bibliography{literature.bib}

\label{lastpage}

\end{document}